\documentclass[twocolumn,showpacs,amsmath,prl]{revtex4}
\usepackage{graphicx}
\usepackage{dcolumn}
\usepackage{amssymb}
\usepackage{bm}
\renewcommand{\d}{{\rm d}}
\renewcommand{\i}{{\rm i}}
\newcommand{\e}{{\rm e}}
\newcommand{\Tr}{{\rm Tr}\;}
\newcommand{\tr}{{\rm tr}\;}
\renewcommand{\Re}{{\rm Re}\;}
\renewcommand{\Im}{{\rm Im}\;}
\newcommand{\sgn}{{\rm sgn}}
\renewcommand{\arctan}{{\rm Tan}^{-1}}
\begin{document}
\title{Orbital Magnetism and Transport Phenomena in Two Dimensional
Dirac Fermions in Weak Magnetic Field}
\author{Masaaki Nakamura}
\affiliation{Department of Applied Physics, Faculty of Science, Science
University of Tokyo, Kagurazaka, Shinjuku-ku, Tokyo 162-8601, Japan}
\date{\today}
\begin{abstract}
 We discuss the orbital magnetism and the Hall effect in the weak
 magnetic field in two dimensional Dirac fermion systems with energy
 gap.  This model is related to the graphene sheet, organic conductors,
 and $d$-density wave superconductors. We found the strong diamagnetism
 and finite Hall conductivity even in gapped systems.  We also discuss
 the relation between the weak-magnetic field formalism and the Landau
 quantization with the Euler-Maclaurin formula.
\end{abstract}

\maketitle


{\em Introduction}---
Experimental studies of graphene have revealed interesting physical
properties\cite{Novoselov,Zhang}. An anomalous quantum Hall effect is
observed where the quantized conductance is described by
$\sigma_{xy}=\pm\frac{e^2}{h}(4n+2)$ ($n=0,1,2,\cdots$).  This result is
explained by the band structure of graphene which has gapless point due
to the hexagonal lattice and a linear dispersion around the gapless
point\cite{Zheng-A,Gusynin-S_2005b,Gusynin-S_2006}.  Moreover, the
minimal conductivity $\sigma_{\rm min}=e^2/h$ par channel is observed,
but theoretical explanation for this is not successful
\cite{Shon-A,Gusynin-S_2005b,Gusynin-S_2006,Ziegler_2007}.  If the Fermi
energy is near the gapless point, the low-energy property of this system
is described by the two-dimensional (2D) Dirac fermions.  Similar
gapless system is also found in the organic conductor
$\alpha$-(BEDT-TTF)$_2$I$_3$, which has a tilted linear dispersion
\cite{Katayama-K-S,Kobayashi-K-S-F}. Moreover, the high-$T_{\rm c}$
superconductor with $d$-density-wave gap is also discussed as a Dirac
fermion system\cite{Yang-N,Sharapov-G-B}.

The physical properties of Dirac fermions in magnetic fields have been
studied by assuming the Landau quantization.  On the other hand, quite
recently, Fukuyama discussed the magnetic susceptibility and the Hall
effect of the gapless Dirac system using the formalism of the weak
magnetic field\cite{Fukuyama_2007}. It is predicted that this system
shows a strong diamagnetism.  On the other hand, gapped systems are also
interesting. Opening of a energy gap is also studied experimentally.
The physical meaning of gap is directly related to the electron density
imbalance between the two sublattices of the bipartite hexagonal lattice
of graphene.
We are also interested in the relation between the results of the theory
of the strong magnetic field where the Landau quantization is essential
and the weak magnetic field, especially for the magnetic susceptibility.
In this paper, we generalize the Fukuyama's work\cite{Fukuyama_2007} to
the gapped systems, and discuss the relation to the strong magnetic
field.

The single particle Hamiltonian of the 2D Dirac fermion is given by
\begin{equation}
 \hat{\cal H}_0=v(p_x\sigma_x+p_y\sigma_y)+\sigma_z\Delta,
  \label{Ham_Dirac.1}
\end{equation}
where $v$ is the Fermi velocity, $\Delta$ is the energy gap, and
$\sigma_i$ ($i=x,y,z$) is the Pauli matrix.  In the presence of the
magnetic field, the momentum operator $\bm{p}=-\i\hbar\nabla$ in the
Hamiltonian is replaced by $\bm{\Pi}=\bm{p}+e\bm{A}/c$ where
$\nabla\times\bm{A}=(0,0,B)$. The Hamiltonian in the many body system is
given by the field operator
$\Psi^{\dag}(\bm{r})=[\psi_{+}^{\dag}(\bm{r}),\psi_{-}^{\dag}(\bm{r})]$,
\begin{equation}
 {\cal H}=\int\d^2\bm{r}\Psi^{\dag}(\bm{r})
 \hat{\mathcal{H}}_0\Psi(\bm{r}).
 \label{Ham_Dirac.2}
\end{equation}

\begin{figure}
 \begin{center}
  \includegraphics[width=75mm]{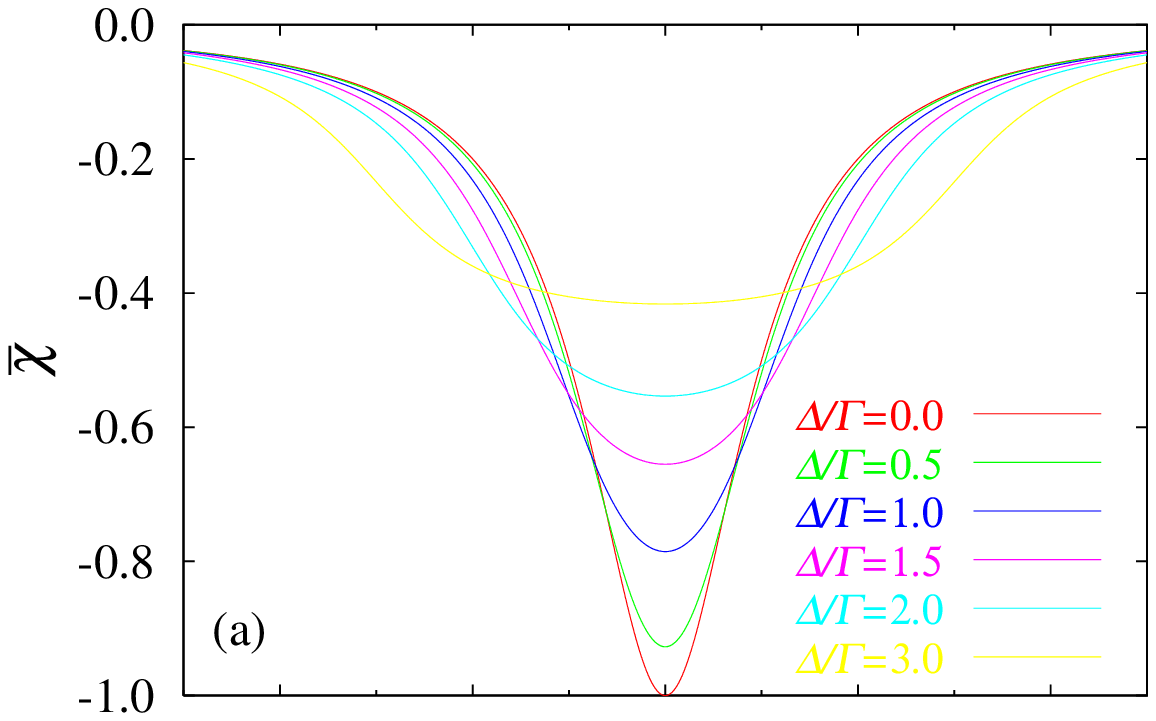}\\
  \includegraphics[width=75mm]{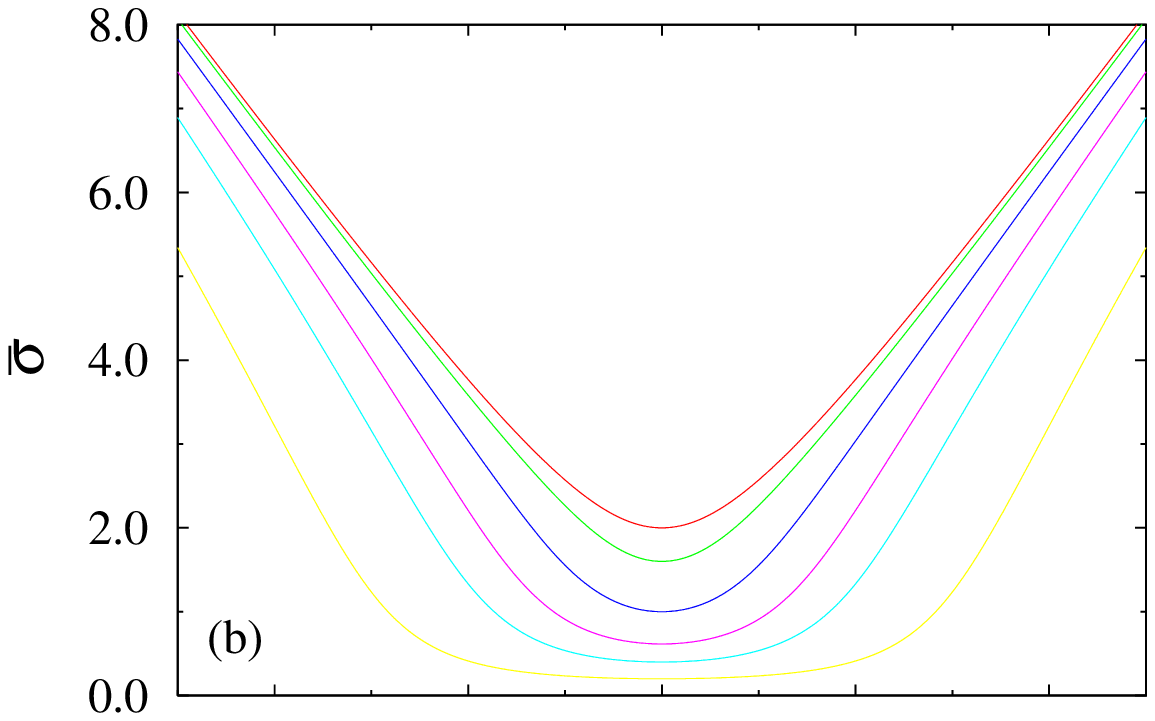}\\  
  \includegraphics[width=75mm]{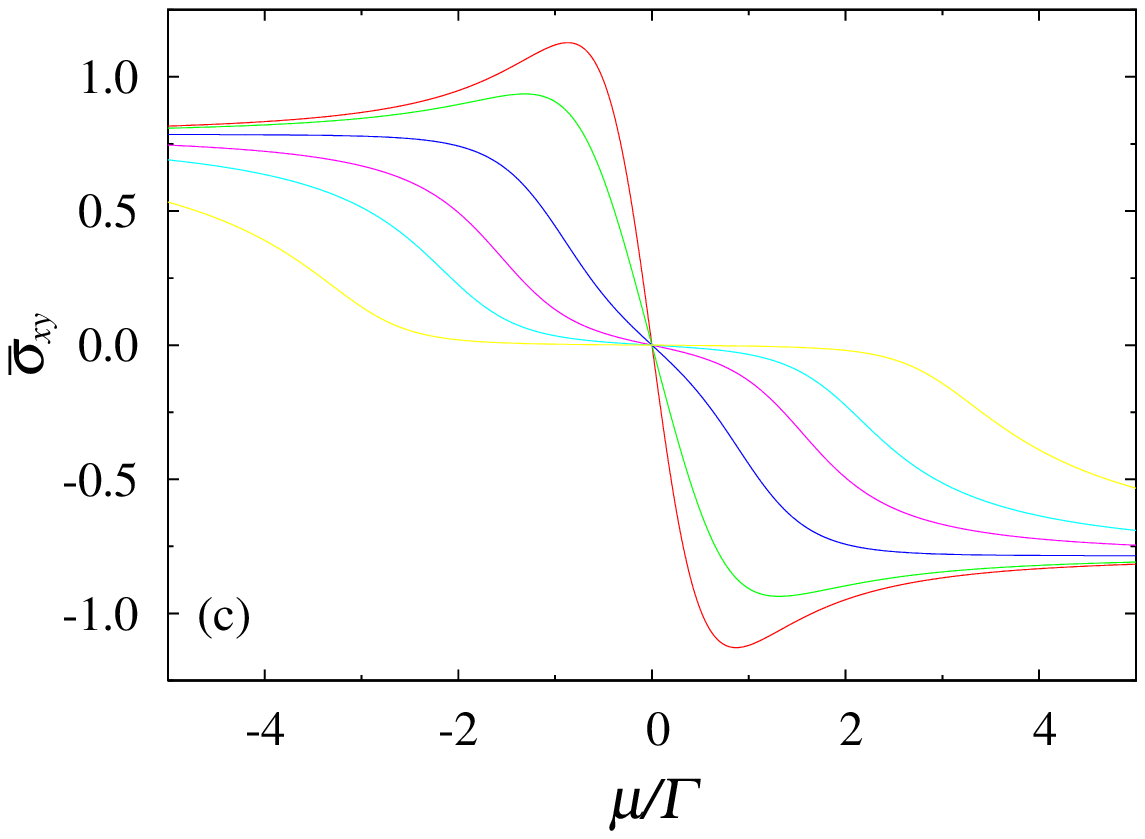}
 \end{center}
 \caption{(a) Magnetic susceptibility $\bar{\chi}\equiv\chi
 3\pi^2c^2\hbar^2/e^2v^2$, (b) Conductivity $\bar{\sigma}\equiv \sigma
 2\pi h/e^2$, and (c) Hall conductivity $\bar{\sigma}_{xy}\equiv
 \sigma_{xy} 4\pi^2 c/e^3v^2B$ of the 2D Dirac fermions in the
 weak-magnetic field, as functions of the scaled chemical potential
 $\mu/\Gamma$. The results (a), (c) of gapless case ($\Delta=0$) are
 derived in Ref.~\cite{Fukuyama_2007}.}  \label{fig.1}
\end{figure}

{\em Orbital Magnetism}---
The general formula of the magnetic susceptibility of the interband
system in terms of the temperature Green function is derived by
Fukuyma\cite{Fukuyama_1970}. This result is obtained by the
Luttinger-Kohn representation\cite{Luttinger-K} for the basic functions
and the Fourier expansion of the vector potential
$\bm{A}(\bm{r})=-\i\bm{A}_{\bm{q}}
(\e^{\i\bm{q}\cdot\bm{r}}-\e^{-\i\bm{q}\cdot\bm{r}})$ with
$B=q_xA_{\bm{q}}^y-q_yA_{\bm{q}}^x$ as,
\begin{equation}
 \chi=\frac{e^2v^4}{\beta V c^2\hbar^2}
  \sum_{n}\sum_{\bm{k}}
  \tr[{\cal G}\sigma_x{\cal G}\sigma_y{\cal G}\sigma_x{\cal G}\sigma_y],
  \label{form_of_susceptibility}
\end{equation}
where $\beta=1/k_{\rm B}T$ and $V$ are the inverse temperature and the
volume of the system, respectively.  The temperature Green function is
given by ${\cal G}\equiv{\cal G}(\bm{k},\i\omega_n)=
(\i\tilde{\omega}_{n}-\hat{\cal H}_0/\hbar)^{-1}$, where
$\i\tilde{\omega}_{n}\equiv\i\omega_n+[\mu+\i\sgn(\omega_n)\Gamma]/\hbar$,
with $\omega_n\equiv(2n+1)\pi/\beta\hbar$ being the Matsubara frequency
of fermions.  Here, we have introduced the scattering rate,
$\Gamma=-\Im\Sigma^{\rm R}$ neglecting the frequency dependence, for
simplicity.  The matrix trace $\tr[\cdots]$ of
eq.~(\ref{form_of_susceptibility}) is calculated as
\begin{equation}
 \frac{16 v^4 k_x^2 k_y^2
  }{[(\i\tilde{\omega}_n)^2-v^2k^2-\tilde{\Delta}^2]^4}
  -\frac{2}{[(\i\tilde{\omega}_n)^2-v^2k^2-\tilde{\Delta}^2]^2},
  \label{sus_matrix_trace}
\end{equation}
where $\tilde{\Delta}\equiv\Delta/\hbar$.  Integrating out the wave
number $\bm{k}$, we have
\begin{equation}
 \frac{1}{V}\sum_{\bm{k}}[\mbox{eq.~(\ref{sus_matrix_trace})}]
  =\frac{1}{3\pi v^2[(\i\tilde{\omega}_n)^2-\tilde{\Delta}^2]}
 \equiv F_1(\i\tilde{\omega}_n).
 \label{sus_k_integrated}
\end{equation}
Finally, taking the $\omega_n$-summation as the contour integration, the
susceptibility is obtained as
\begin{align}
 \chi
 =&\frac{e^2v^4}{c^2\hbar}\frac{1}{\beta\hbar}\sum_n
 F_1(\i\tilde{\omega}_n)\\
 =&\frac{e^2v^4}{c^2\hbar}\frac{-1}{2\pi\i}
 \int_{-\infty}^{\infty}\d\omega f(\hbar\omega)
 \left[F_1(\omega+\i\tilde{\Gamma})-F_1(\omega-\i\tilde{\Gamma})\right],
 \label{sus_formula.2}
\end{align}
where $f(x)\equiv(\e^{\beta(x-\mu)}+1)^{-1}$ is the Fermi distribution
function and $\tilde{\Gamma}\equiv\Gamma/\hbar$.  Especially, in the
zero-temperature limit $f(x)\to\theta(\mu-x)$, we have
\begin{align}
 \chi =&-\frac{e^2v^2}{6\pi^2c^2\hbar^2\Delta}
 \left[\arctan\frac{\mu+\Delta}{\Gamma}
 -\arctan\frac{\mu-\Delta}{\Gamma}\right]\\
 =&-\frac{e^2v^2}{3\pi^2c^2\hbar^2}\frac{\Gamma}{\Gamma^2+\mu^2} \qquad
 (\Delta\to 0).
\end{align}
In the $\Delta\to 0$ limit, this result coincides with that of the
gapless system\cite{Fukuyama_2007}, as expected. Moreover in the clean
limit $\Gamma\to 0$, this gives the $\delta$-function with negative
sign.  The chemical potential dependence of the susceptibility is shown
in Fig.~\ref{fig.1}(a). This result shows that a strong diamagnetism
appears even in the presence of a gap, and the susceptibility takes the
minimum value when $\mu$ is in the middle of the gap. This result is
similar to the case of bismuth\cite{Fukuyama_1970}.

Now we will show that the magnetic susceptibility (\ref{sus_formula.2})
is also obtained by the Landau quantization formalism.  To do this, it
is sufficient to derive eq.~(\ref{sus_k_integrated}).  The Hamiltonian
(\ref{Ham_Dirac.1}) in the magnetic field is given by
\begin{equation}
\hat{\cal H}_0
  =\left[\begin{array}{cc}
   \Delta & v\Pi_- \\ v\Pi_+ & -\Delta
	\end{array}\right],
  \label{Ham_Dirac.3}
\end{equation}
where $\Pi_{\pm}\equiv\Pi_x\pm\i\Pi_y$.  Since the commutation relation
between these operators is $[\Pi_{\pm},\Pi_{\mp}]=\mp 2eB\hbar/c$, there
is the correspondence to the creation and the annihilation operators of
the harmonic oscillator: $\Pi_{+}\to\sqrt{2}\frac{\hbar}{l}a^{\dag}$,
$\Pi_{-}\to\sqrt{2}\frac{\hbar}{l}a$ where $l\equiv\sqrt{c\hbar/eB}$ and
$eB>0$. Then the eigenvalue and the eigenstate of (\ref{Ham_Dirac.3}) is
obtained as
\begin{align}
 &
 \hat{\cal H}_0^2|k\rangle\rangle=M_k^2|k\rangle\rangle,\quad
  M_k\equiv\sqrt{2k\frac{\hbar^2v^2}{l^2}+\Delta^2},\\
 &
 |k\rangle\rangle
 \equiv\frac{1}{\sqrt{2}}
 \left[\begin{array}{c}
  |k-1\rangle \\ |k\rangle
       \end{array}\right] \quad(k>0),\quad
 |0\rangle\rangle
 \equiv
 \left[\begin{array}{c} 0 \\ |0\rangle\end{array}\right],
 \label{2D_Dirac_eigenstate.2}
\end{align}
where $k=0,1,2,3,\cdots$ denote the Landau levels with positive energy,
and $|k\rangle$ corresponds to the number state of the harmonic
oscillator.  Note that $k=0$ state is special: one component has zero
amplitude, and the normalization factor is different from $k>0$ cases.

According to the functional integral method, the thermodynamic
potential is given by the temperature Green function as
\begin{align}
 \Omega(B)
 =&-\frac{1}{\beta\hbar}\sum_{n=-\infty}^{\infty}
 \Tr\log(-\i\tilde{\omega}_n+\hat{\cal H}_0/\hbar)\nonumber\\
 =&-\frac{2V}{\pi l^2\beta\hbar}
 \sum_{n=-\infty}^{\infty}
 \biggl[
 \sum_{k=1}^{\infty}
 \log\left|(\i\tilde{\omega}_n)^2-(M_{k}/\hbar)^2\right|\nonumber\\
 &
 +\frac{1}{2}
 \log\left|(\i\tilde{\omega}_n)^2-(M_0/\hbar)^2\right|
 \biggr].
 \label{sus_potential.3}
\end{align}
Here factor $V/\pi l^2$ in eq.~(\ref{sus_potential.3}) is the degeneracy
of the Landau levels for $|k|>0$.  For $k=0$, the degeneracy is $V/2\pi
l^2$. Using the Euler-Maclaurin formula,
\begin{equation}
 \frac{1}{2}g(0)+
 \sum_{k=1}^{\infty}g(k)
  \simeq\int_0^{\infty} g(x)\,\d x-\frac{1}{12}g'(0),
\end{equation}
the summation over the index of the Landau levels in
eq.~(\ref{sus_potential.3}) is evaluated in the $B\to 0$ limit as
\begin{align}
 \frac{1}{\pi l^2}
 \left[\cdots\right]
 =&
 \frac{1}{\pi l^2}
 \int_0^{\infty}\d x
 \log\left|(\i\tilde{\omega}_n)^2-2xv^2/l^2-\tilde{\Delta}^2\right|
 \nonumber\\
 &
 +\frac{1}{6\pi l^4}
 \frac{v^2}{(\i\tilde{\omega}_n)^2-\tilde{\Delta}^2}.
 \label{Euler-Maclaurin.3}
\end{align}
Since the first term of eq.~(\ref{Euler-Maclaurin.3}) is constant with
respect to the magnetic field, the contribution to the magnetic
susceptibility is given as
\begin{equation}
 -\frac{1}{2V}
 \left(\frac{c\hbar}{ev^2}\right)^2
 \frac{\partial^2\Omega}{\partial B^2}
  =
  \frac{1}{3\pi v^2}
  \sum_{n=-\infty}^{\infty}
  \frac{1}{(\i\tilde{\omega}_n)^2-\tilde{\Delta}^2}.
\end{equation}
Thus eq.~(\ref{sus_k_integrated}) is obtained. This means that the two
different formalisms for the magnetic susceptibilities give the same
result.

\bigskip

{\em Hall effect}---
Next, we consider the transport properties of this system.  The
conductivity of the 2D Dirac system is already derived in
Ref.~\cite{Sharapov-G-B}.  To clarify the derivation of the Hall
conductivity $\sigma_{xy}$, we rederive this result first. The
conductivity is given by the Kubo formula,
\begin{equation}
 \Re\sigma_{\mu\nu}
  =\lim_{\omega\to 0}
  \frac{\Im\tilde{\Pi}_{\mu\nu}(\bm{0},\omega+\i\eta)}{\hbar\omega},
   \label{Kubo_form}
\end{equation}
where $\tilde{\Pi}_{\mu\nu}(\bm{q},\omega)\equiv
\Pi_{\mu\nu}(\bm{q},\omega)-\Pi_{\mu\nu}(\bm{q},0)$.  The polarization
function in the Matsubara form is given by
\begin{align}
 &
 \Pi_{\mu\nu}(\bm{q},\i\nu_m)
   =\frac{1}{V}\int_{0}^{\beta\hbar}\d\tau\e^{\i\nu_m\tau}
 \langle{\cal T}_{\tau}
 J_{\mu}(\bm{q},\tau)J_{\nu}(\bm{0},0)\rangle\label{Pi_CC},\\
 &
 \tilde{\Pi}_{\mu\nu}(\bm{0},\i\nu_m)=
 -\frac{e^2v^2}{\beta\hbar V}
 \sum_{\bm{k}}\sum_{n}
 \tr\left[{\cal G}\sigma_{\mu}{\cal G}_+\sigma_{\nu}\right],
 \label{Pi_CC2}
\end{align}
where ${\cal G}_+\equiv{\cal G}(\bm{k},\omega_n+\nu_m)$ with $\nu_m=
2\pi m/\beta\hbar$ being the Matsubara frequency of bosons. The current
operator is given by
\begin{equation}
 J_{\mu}(\bm{q},\tau)=
  -ev\sum_{\bm{k}}
  \Psi^{\dag}(\bm{k}^+,\tau)\sigma_{\mu}\Psi(\bm{k}^-,\tau),
\end{equation}
where $\bm{k}^{\pm}\equiv\bm{k}\pm\bm{q}/2$.  The matrix trace
$\tr[\cdots]$ in eq.~(\ref{Pi_CC2}) for $\mu=\nu=x$ is calculated
as
\begin{equation}
 \frac{2[(\i\tilde{\omega}_n)(\i\tilde{\omega}_n^+)-\tilde{\Delta}^2]
 }{[(\i\tilde{\omega}_n)^2-v^2k^2-\tilde{\Delta}^2]
 [(\i\tilde{\omega}_n^+)^2-v^2k^2-\tilde{\Delta}^2]}
 \equiv F_2(\i\tilde{\omega}_n,\i\tilde{\omega}_n^+),
 \label{sxx_matrix_trace}
\end{equation}
where $\omega_n^+\equiv\omega_n+\nu_m$.  The summation of the Matsubara
frequency is carried out in the following way,
\begin{align}
 \lefteqn{\frac{1}{\beta\hbar}
 \sum_n F_i(\i\tilde{\omega}_n,\i\tilde{\omega}_n^+)
 =\frac{-1}{2\pi\i}\int_{-\infty}^{\infty}\d \omega
 f(\hbar\omega)} \label{Matsubara_sum.2}\\
 &\times
 [F_i(\omega+\i\tilde{\Gamma},\omega+\i\nu_m+\i\tilde{\Gamma})
 -F_i(\omega-\i\tilde{\Gamma},\omega+\i\nu_m+\i\tilde{\Gamma})
 \nonumber\\
 &
 +F_i(\omega-\i\nu_m-\i\tilde{\Gamma},\omega+\i\tilde{\Gamma})
 -F_i(\omega-\i\nu_m-\i\tilde{\Gamma},\omega-\i\tilde{\Gamma})].
\nonumber
\end{align}
After expanding eq.~(\ref{Matsubara_sum.2}) in the linear order of
$\i\nu_m$ and the partial integration of $\omega$, it follows from
eqs.~(\ref{Kubo_form}) and (\ref{Pi_CC2}) that the conductivity is
calculated as
\begin{align}
 &
 \sigma
 =\frac{e^2}{4\pi^2}
 \int_{-\infty}^{\infty}\d\omega [-f'(\hbar\omega)]
 {\cal A}_2(\hbar\omega,\Gamma,\Delta),\\
 &
 {\cal A}_2(x,\Gamma,\Delta)\equiv\nonumber\\
 &
 \frac{\Gamma^2-\Delta^2+x^2}{2\Gamma|x|}
 \left(\frac{\pi}{2}-
 \arctan\frac{\Delta^2+\Gamma^2-x^2}{2\Gamma|x|}\right)+1.
  \label{sigma_xx.3}
\end{align}
Here the last term in eq.~(\ref{sigma_xx.3}) stems from the first and
forth terms in eq.~(\ref{Matsubara_sum.2}) where two frequencies have
the same analytic properties, while the other terms originate from the
second and the third terms.  At zero temperature,
$f'(x)=-\delta(x-\mu)$, we have $\sigma=\frac{e^2}{2\pi h} {\cal
A}_2(\mu,\Gamma,\Delta)$ which is derived in Ref.~\cite{Sharapov-G-B}.
The chemical potential dependence of the conductivity is shown in
Fig.~\ref{fig.1}(b).  In the limit $\mu/\Gamma\to 0$, we have the
minimal conductivity, $\sigma_{\rm min}=\frac{e^2}{\pi
h}(1+\Delta^2/\Gamma^2)^{-1}$.  The dependence of the magnetic field $B$
can be calculated in the similar way of $\sigma_{xy}$ which is discussed
below [eq.~(\ref{Pi_CC3}) for $(\mu,\nu)=(x,x)$]. One can show, however,
that the conductivity has no $B$-dependence in the linear order.

The Hall conductivity $\sigma_{xy}$ in a weak magnetic field is obtained
in the following way\cite{Fukuyama-E-W,Fukuyama_2006}: The Fourier
expansion of the vector potential is chosen as
$\bm{A}(\bm{r})=\bm{A}_{\bm{q}}\e^{\i\bm{q}\cdot\bm{r}}$.  Then by the
perturbative expansion of eq.~(\ref{Pi_CC}) in terms of the Hamiltonian
in the magnetic field ${\cal H}-\bm{A}_{\bm{q}}\cdot\bm{J}(-\bm{q})/c$,
$\sigma_{xy}$ is given by the three point correlation function, and the
linear term of the magnetic field $B=\i(q_x A_{\bm{q}}^{y}-q_y
A_{\bm{q}}^{x})$ is obtained by the $\bm{q}$ expansion of the Green
function $\partial_{k_{\mu}}{\cal G}=v{\cal G}\sigma_{\mu}{\cal G}$ as,
\begin{align}
 &
 \Pi_{\mu\nu}(\bm{q},\i\nu_m)=
 -\sum_{\alpha=x,y}\frac{A_{\bm{q}\alpha}}{c\hbar}\frac{1}{V}
 \int_{0}^{\beta\hbar}\d\tau
 \int_{0}^{\beta\hbar}\d\tau'\nonumber\\
 &\times
 \e^{\i\nu_m\tau}
 \langle{\cal T}_{\tau}
 J_{\mu}(\bm{q},\tau)J_{\alpha}(-\bm{q},\tau')J_{\nu}(\bm{0},0)\rangle
 \label{Pi_CC3},\\
 &
 \tilde{\Pi}_{\mu\nu}(\bm{0},\i\nu_m)
 =-\i\frac{B}{c\hbar}
 \frac{e^3v^4}{2}
 \frac{1}{V\beta\hbar}
 \sum_{\bm{k},n}\nonumber\\
 &\times
 \tr[
 \sigma_{\mu}{\cal G}_+\sigma_x{\cal G}_+\sigma_{\nu}{\cal G}\sigma_y{\cal G}
 -\sigma_{\mu}{\cal G}_+\sigma_y{\cal G}_+\sigma_{\nu}{\cal G}\sigma_x{\cal G}
 \nonumber\\
 &
+\sigma_{\mu} {\cal G}_+\sigma_x {\cal G}_+\sigma_y {\cal G}_+\sigma_{\nu}{\cal G}
-\sigma_{\mu} {\cal G}_+\sigma_y {\cal G}_+\sigma_x {\cal G}_+\sigma_{\nu} {\cal G}
 \nonumber\\
 &
+
 \sigma_{\mu}{\cal G}_+\sigma_{\nu}{\cal G}\sigma_x{\cal G}\sigma_y{\cal G}
-
 \sigma_{\mu}{\cal G}_+\sigma_{\nu}{\cal G}\sigma_y {\cal G}\sigma_x{\cal G}
 ].\label{Pi_CC4}
\end{align}
The matrix trace $\tr[\cdots]$ for $(\mu,\nu)=(x,y)$ in
eq.~(\ref{Pi_CC4}) becomes
\begin{equation}
 \frac{-4[(\i\tilde{\omega}_n)^2-(\i\tilde{\omega}_n^+)^2]
  [(\i\tilde{\omega}_n)(\i\tilde{\omega}_n^+)-\tilde{\Delta}^2]
 }{[(\i\tilde{\omega}_n)^2-v^2k^2-\tilde{\Delta}^2]^2
 [(\i\tilde{\omega}_n^+)^2-v^2k^2-\tilde{\Delta}^2]^2}
 \equiv F_3(\i\tilde{\omega}_n,\i\tilde{\omega}_n^+).
 \label{dirac_si_xy_trace}
\end{equation}
Using eq.~(\ref{Matsubara_sum.2}), the Hall conductivity is obtained as
in the same way of the conductivity $\sigma$:
\begin{align}
 &
 \sigma_{xy}=-\frac{e^3v^2B\hbar}{4\pi^2 c}
 \int_{-\infty}^{\infty}\d \omega [-f'(\hbar\omega)]
 {\cal A}_3(\hbar\omega,\Gamma,\Delta),\\
 &
 {\cal A}_3(x,\Gamma,\Delta)\equiv
 \frac{\Gamma^2-\Delta^2+x^2}{4\Gamma^2x|x|}
 \left(\frac{\pi}{2}-
 \arctan\frac{\Delta^2+\Gamma^2-x^2}{2\Gamma|x|}\right)\nonumber\\
 &+   
 \frac{1}{(\Gamma^2+x^2+\Delta^2)^2-4\Delta^2x^2}
 \left(
 \frac{(\Delta^2-x^2)^2-\Gamma^4}{2\Gamma x}
 +\frac{4\Gamma x}{3}
 \right),
\end{align}
where the last term of ${\cal A}_3$ which is proportional to
$\frac{4\Gamma x}{3}$ stems from the first and the forth terms in
eq.~(\ref{Matsubara_sum.2}).  At zero temperature, we have
$\sigma_{xy}=-\frac{e^3v^2B}{4\pi^2c} {\cal A}_3(\mu,\Gamma,\Delta)$.
The chemical potential dependence of the Hall conductivity is shown in
Fig.~\ref{fig.1}(c).  In the dirty limit $\Gamma\to\infty$, and
$\Delta,\mu\to 0$, the Drude-Zener formula is obtained with
$\Gamma=\hbar/2\tau$, where $\tau$ is the mean-free time of
qusiparticles.  This is consistent with the result obtained by formalism
of the Landau quantization\cite{Gusynin-S_2006}.


\bigskip

{\em Thermal Hall effect}---
Finally, we consider the thermal conductivity. The thermal transport of
the Dirac system is discussed in the same way of
Refs.\cite{Gusynin-S_2005a,Sharapov-G-B,Ferrer-G-I}: For the Hamiltonian
density
\begin{equation}
 \varepsilon(\bm{r})\equiv
  \Psi^{\dag}(\bm{r})\hat{\mathcal{H}}_0\Psi(\bm{r}),
\end{equation}
where the differentiation operator in the one particle Hamiltonian
(\ref{Ham_Dirac.1}) is replaced as
$\partial_{\alpha}\to(\partial_{\alpha}
-{\displaystyle\mathop{\partial_{\alpha}}^{\leftarrow}})/2$, by the
partial integration of eq.~(\ref{Ham_Dirac.2}).  Then the time
derivative of the local energy is given by
\begin{equation}
 \dot{\varepsilon}
   =\dot{\Psi}^{\dag}\hat{\mathcal{H}}_0\Psi
   +\Psi^{\dag}\hat{\mathcal{H}}_0\dot{\Psi}.
\end{equation}
The energy current $\bm{J}_E$ is determined so that it satisfies the
following continuum equation,
\begin{equation}
 \dot{\varepsilon}+\nabla\cdot\bm{J}_E=0.
  \label{th_current_continuum}
\end{equation}
Thus we obtain
\begin{equation}
 J_E^{\alpha}
  =\i\frac{v}{2}(\Psi^{\dag}\sigma_{\alpha}\dot{\Psi}-
  \dot{\Psi}^{\dag}\sigma_{\alpha}\Psi).
\end{equation}
The current-energy ($\Pi_{\mu\nu}^{CE}$) and the energy-energy
($\Pi_{\mu\nu}^{EE}$) correlation functions are defined as in the same
way of eq.~(\ref{Pi_CC}). These correlation functions have similar
structure of the current-current correlation function:
\begin{align}
 &
 \lim_{\omega\to 0}
 \frac{\beta\Im\tilde{\Pi}_{xy}^{CE}(\bm{0},\omega+\i\eta)}{\hbar\omega}
 \nonumber\\
 &
 =-\frac{e^2v^2B}{4\pi^2 c}\int_{-\infty}^{\infty}\d x
 f'(x)[\beta(x-\mu)]{\cal A}_3(x,\Gamma,\Delta),\label{Im_Pi_CE}\\
 &
 \lim_{\omega\to 0}
 \frac{\beta^2\Im\tilde{\Pi}_{xy}^{EE}(\bm{0},\omega+\i\eta)}{\hbar\omega}
 \nonumber\\
 &
 =\frac{ev^2B}{4\pi^2 c}\int_{-\infty}^{\infty}\d x
 f'(x)[\beta(x-\mu)]^2{\cal A}_3(x,\Gamma,\Delta).
 \label{Im_Pi_EE}
\end{align}
In the zero temperature limit, using the relation $\int_{-\infty}^\infty
f'(x)[\beta(x-\mu)]^n\d x=1$ ($n=0$), $0$ ($n=\mbox{odd}$), $2
n!(1-2^{-n+1})\zeta(n)$ ($n=\mbox{even}\geq2$), it turns out that
eq.~(\ref{Im_Pi_CE})$=0$ and eq.~(\ref{Im_Pi_EE})$=-
\frac{\pi^2}{3}\frac{ev^2B}{4\pi^2c} {\cal A}_3(\mu,\Gamma,\Delta) $
where $\zeta(2)=\pi^2/6$ is used.  Since the thermopower related to
$\Pi^{CE}$ does not contribute to the thermal Hall conductivity
$\kappa_{xy}$, the Wiedemann-Franz law
$\kappa_{xy}/\sigma_{xy}T=(\pi^2/3)(k_{\rm B}/e)^2$ is satisfied for the
Hall and the thermal Hall conductivities in the low temperature limit.

\bigskip

{\em Summary}---
We have discussed the magnetic susceptibility, the Hall conductivity and
the thermal Hall conductivity of the 2D Dirac fermions in the weak
magnetic field.  We have shown that strong diamagnetism appears even in
the gapped system if the Fermi energy is in the middle of the gap.  We
have also shown that the magnetic susceptibility derived by the weak
magnetic field formalism is equivalent with that of the Landau
quantization with the Euler-Maclaurin formula.  The other results
including the Hall conductivity are also consistent with those of the
Landau quantization formalism.

\bigskip

{\em Acknowledgment}---
The author is grateful to Hidetoshi Fukuyama for many helpful guidances
and fruitful discussions.  He also thanks Shunsuke Furukawa for
discussions.


\begin{references}

 \bibitem{Novoselov}
 K.~S.~Novoselov {\it et al.}, Nature (London) {\bf 438}, 197 (2005).
 
 \bibitem{Zhang}
 Y.~Zhang {\it et al.}, Nature (London) {\bf 438}, 201 (2005).

 \bibitem{Zheng-A}
 Y.~Zheng and T.~Ando, Phys. Rev. B {\bf 65}, 245420 (2002).

 \bibitem{Gusynin-S_2005b}
 V. P. Gusynin and S. G. Sharapov,
 Phys. Rev. Lett. {\bf 95}, 146801 (2005).
 
 \bibitem{Gusynin-S_2006}
 V. P. Gusynin and S. G. Sharapov,
 Phys. Rev. B {\bf 73}, 245411 (2006)

 \bibitem{Shon-A}
 N.~H.~Shon and T.~Ando,
 J.~Phys.~Soc.~Jpn. {\bf 67}, 2421 (1998).

 \bibitem{Ziegler_2007}
 K. Ziegler, cond-mat/0701300, and references therein.
 
 \bibitem{Katayama-K-S}
 S.~Katayama, A.~Kobayashi, and Y.~Suzumura,
 J. Phys. Soc. Jpn {\bf 75}, 054705 (2006).

 \bibitem{Kobayashi-K-S-F}
 A.~Kobayashi, S.~Katayama, Y.~Suzumura, and H.~Fukuyama,
 to appear in J. Phys. Soc. Jpn.

 \bibitem{Yang-N}
 X. Yang and C. Nayak,
 Phys. Rev. B {\bf 65}, 064523 (2002).

 \bibitem{Sharapov-G-B}
 S. G. Sharapov, V. P. Gusynin, and H. Beck,
 Phys. Rev. B {\bf 67}, 144509 (2003).

 \bibitem{Fukuyama_2007}
 H. Fukuyama, cond-mat/0703010, to appear in J. Phys. Soc. Jpn.
 
 \bibitem{Fukuyama_1970}
 H. Fukuyama, Prog. Theor. Phys. {\bf 45}, 704 (1971).
 
 \bibitem{Luttinger-K}
 J.~M.~Luttinger and W.~Kohn,
 Phys. Rev. {\bf 97}, 869 (1955).

 \bibitem{Fukuyama-E-W}
 H. Fukuyama, H. Ebisawa, and Y. Wada,
 Prog. Theor. Phys. {\bf 42}, 494 (1969).
 
 \bibitem{Fukuyama_2006}
 H. Fukuyama, Ann. Phys. {\bf 15}, 520 (2006).
 

 \bibitem{Gusynin-S_2005a}
 V. P. Gusynin and S. G. Sharapov,
 Phys. Rev. B {\bf 71}, 125124 (2005).

 \bibitem{Ferrer-G-I}
 E. J. Ferrer, V. P. Gusynin, and V. de la Incera,
 Eur. Phys. J. B {\bf 33}, 397 (2003).
 
 
\end{references}
\end{document}